\documentclass[journal=jacsat,manuscript=article]{achemso}

\setkeys{acs}{maxauthors = 0}

\usepackage{chemformula} 
\usepackage[T1]{fontenc} 

\usepackage{xr}

\usepackage{caption}
\usepackage{subcaption}

\usepackage{amssymb,amsmath}
\usepackage{soul}
\usepackage{rotating}
\usepackage{chemscheme} 
\usepackage{chemfig} 
\usepackage{makecell} 
\usepackage{multirow}

\author{D. Thirumalai}
\email{dave.thirumalai@gmail.com}
\affiliation{Department of Chemistry, The University of Texas at Austin, Austin, TX 78712}
\alsoaffiliation{Department of Physics, The University of Texas at Austin, Austin, TX 78712}
\author{Abhinaw Kumar}
\affiliation{Department of Chemistry, The University of Texas at Austin, Austin, TX 78712}
\author{Debayan Chakraborty}
\affiliation{Department of Chemistry, The University of Texas at Austin, Austin, TX 78712}
\author{John E. Straub}
\affiliation{Department of Chemistry, Boston University, Boston, MA 78712}
\author{Mauro L. Mugnai}
\affiliation{Institute of Soft Matter Synthesis and Metrology, Georgetown University, Washington, DC 20057}

\title{Conformational Fluctuations and Phases in Fused in Sarcoma (FUS) Low-Complexity Domain}

\abbreviations{}

\begin{document}







\clearpage
\begin{abstract}
The well known phenomenon of phase separation in synthetic polymers and proteins has become a major topic in biophysics because it  has  been invoked as a mechanism of compartment formation in cells, without the need for membranes. Most of the coacervates (or condensates) are composed of Intrinsically Disordered Proteins (IDPs) or regions that are structureless, often in interaction with RNA and DNA. One of the more intriguing IDPs is the 526-residue RNA binding protein, Fused In Sarcoma (FUS), whose monomer conformations and condensates exhibit unusual behavior that are sensitive to solution conditions. By focussing principally on the N-terminus low complexity domain (FUS-LC comprising residues 1-214) and other truncations, we rationalize the findings of solid state NMR experiments, which show that FUS-LC adopts a non-polymorphic  fibril (core-1) involving residues 39-95, flanked by fuzzy coats on both the N- and C- terminal ends. An alternate structure (core-2), whose free energy is comparable to core-1, emerges only in the truncated construct (residues 110-214).  Both core-1 and core-2 fibrils are stabilized by a Tyrosine ladder as well as hydrophilic interactions. The morphologies (gels, fibrils, and glass-like behavior) adopted by FUS seem to vary greatly, depending on the experimental conditions. The effect of phosphorylation is site specific and affects the stability of the fibril depending on the sites that are phosphorylated.   Many of the peculiarities  associated with FUS may also be shared by other IDPs, such as TDP43 and hnRNPA2.  We outline a number of problems for which there is no  clear molecular understanding.
\end{abstract}

\section{Introduction}
Phase separation in polymers dispersed in a solvent has been known for a long time~\cite{FloryBook}. If the interaction between the solvent and the monomers is not favorable,  phase separation occurs  producing  a polymer rich phase that coexists with  polymer poor phase. If this process occurs at equilibrium, it can be described using the venerable mean field Flory-Huggins (FH) theory,\cite{FloryBook,FloryJCP1942,HugginsJCP1942} which becomes increasingly accurate as the degree of polymerization increases, because the mixing entropy due to polymers is $\propto \frac{1}{N}$ ($N$ is the degree of polymerization).  Extension of the FH theory to polyelectrolyte solutions was given by Overbeek and Voorn~\cite{Overbeek57JCellComparativePhys}, who set out to explain the remarkable observations by Bugenberg de Jong, showing complex coacervation (a term coined in 1929~\cite{Jong29ProcKoninkl}) in polyelectrolytes.  The dense phase in a two component polyelectrolyte system (gum arabic, a complex mixture of glycoproteins and polysaccharides,  and gelatin) consisted of the polymers with water, and the dilute phase was a solution containing one or both polymers. Bugenberg de Jong pointed out that biological membranes and the complex coacervates may share common characteristics. Indeed, Overbeek and Voorn suggested potential connections to complex coacervation  in biology in the concluding section of their seminal paper.\cite{Overbeek57JCellComparativePhys}  However, the demonstration that condensate (or equivalently coacervate) formation involving biomolecules  could produce organelles without membranes by {\color{black} a phase separation mechanism, has given a  new perspective} on how the cytosol is organized \cite{Brangwynne2009Science,Banani17NatRevMolCellBiol,Alberti21NatRevMolCellBiol,Hyman14AnnRevCellDevBiol}. {\color{black}A key paper}\cite{ Brangwynne2009Science} that appeared in 2009 has set in motion a growing enterprise that has certainly generated considerable interest.

Many of the proteins that produce cellular condensates are intrinsically disordered or contain regions that do not adopt discernible structures, and are often in complex with RNA.  Because condensate (we will use this modern terminology as opposed to coacervates as it seems to be the accepted description in biophysics and biology) formation  leads to high local density of biomolecules, it could profoundly affect a number of functions, as discussed recently by Alberti and Hyman (see Figure 1 in \cite{Alberti21NatRevMolCellBiol}).  A number of other articles have described the functional implications of condensate formation~\cite{Potter16TrendscellBiol,Holehouse18Biochem}. These and scores of others, have provided great impetus in  developing theories and computational models to describe many aspects of phase separation as well as the mechanical and relaxation properties of  the  condensates. It is hard not to notice \textcolor{black}{that a number of researchers are dedicated in elucidating}  the principles of condensate formation from molecular monomer level to mesoscopic scales.

In this perspective, we focus only on the RNA binding protein, Fused in Sarcoma (FUS), which has become a poster child for IDRs (intrinsically disordered protein that also has ordered regions). The N-terminal region of the 526-residue FUS contains the low complexity domain, FUS-LC, (residues 1-214) which is rich in QGSY (Q = Glutamine, G=Glycine, S=Serine, and {\color{black} Y= Tyrosine}) repeats. Residues beyond 214 contain the RNA binding domain, whose role in creating various morphologies of the condensates are not fully understood. Here, we focus on the effects of sequence and length on the conformations and phases of FUS-LC  and several truncated variants.  We first describe the properties of the monomer ensembles of several constructs of FUS-LC, with particular focus on the sparsely populated excited states, referred to as N$^*$. Interestingly, the N$^*$ states reveal characteristics that are found in the fibrils of FUS-LC\cite{Murray17Cell} and the truncated N-terminal variant.\cite{Lee20NatComm}  We find that effects of phosphorylation are site specific, as noted in the experiments~\cite{Murray17Cell}. Phosphorylation of residues within the fibril region {\color{black} (residues 39-95)} in  FUS-LC has a greater impact on the stability of the fibril than phosphorylation of residues outside the core.  Our results for the FUS monomer have important implications in the early events of condensate formation, as well as the aging of condensates into solid-like aggregates.  Finally,  we also raise a few puzzles  that require additional scrutiny through experimentation as well as well-designed computations. 

\section{Results}


{\bf Sequence Analysis:} The 214-residue FUS-LC is rich in QGSY.  Out of the 214 residues, 174 are from only 4 amino acids (the numbers are: Glutamine, Q = 43, Serine, S = 51, Tyrosine, Y = 27, and Glycine, G = 53).  There are relatively few charged residues (6 in total). Thus, electrostatic interactions are negligible, except when certain residues (S and or Y) are phosphorylated or mutated (G156E)~\cite{Patel15Cell,Berkeley21BJ}.   Because the 1-214 region is abundant in these four residues, with no special pattern along the sequence, FUS-LC is a low complexity sequence, which may quantified by sequence or word entropy. FUS-LC is often likened to prion-domains found in yeast prion proteins, because it is enriched in uncharged polar amino acids, as well as glycine.\cite{GitlerPrion2011,BalohPrion2011}  The basis for picturing FUS-LC as a prion-like domain (PLD) comes from sequence analysis~\cite{Michelitsch00PNAS} and experiments on Sup35 and Ure2p from {\it Saccharomyces Cerevisae}. Among the three major domains of Sup35 (yeast prion), residues 1-123 comprising the N-domain, also referred to as the prion domain, is rich in amino acids N  and Q with very few charged residues. The N-domain alone could form  self-propagating (by the Griffith mechanism~\cite{Griffith67Nature}), inheritable amyloids~\cite{Ter-Avanesyan94Genetics,King04Nature}. Thus, from the perspective of  sequence and the ability to self-propagate, the Sup35 prion domain is different from FUS-LC. Despite the differences in sequence composition between FUS-LC and Sup35 (we do not include mammalian prions whose ground state forms an ordered structure in the C-terminus), the FUS-LC has been christened as a PLD~\cite{Patel15Cell}  - a terminology that we adopt here. 

{\bf The ground states of FUS-LC and FUS-LC-C are random coils:} Like many other IDPs, the ground state of FUS-LC and a number of variants behave as polymers in good solvents (Flory random coils), which from a polymer theory perspective is surprising but it is what it is. \textcolor{black}{Because these sequences contain hydrophobic and polar residues one would expect that there ought to be detectable deviations from random coil  behavior. That this is not so, at least in global properties (radius of gyration for example) is a surprise.}
These findings are consistent with narrow chemical shift dispersions of monomers in solution~\cite{Burke15MolCell} in a truncated variant ( residues 1-163). In particular, the C$_{\alpha}$ and C$_{\beta}$ chemical shifts do not deviate significantly from assignments based on what is expected for known random coil structures. {\color{black} Moreover, it was concluded that the chemical shifts} in the FUS-LC domain does not change significantly even in the full length FUS that includes the RNA binding domain~\cite{Burke15MolCell}. The finding that FUS-LC is disordered is in accord with the random coil behavior in many other IDPs~\cite{Baul19JPCB}. In a number of IDPs, the radius of gyration ($R_g$), as well as the hydrodynamic radius ($R_h$) follow Flory's scaling law,  $R_g \approx a_D N_T^{\nu}$ and $R_h \approx N_T^{\nu}$, where the exponent $\nu \approx 0.58$, and $N_T$ is the protein length~\cite{Baul19JPCB}.



 \begin{figure}[htbp]
\begin{center}
\includegraphics[width=\textwidth]{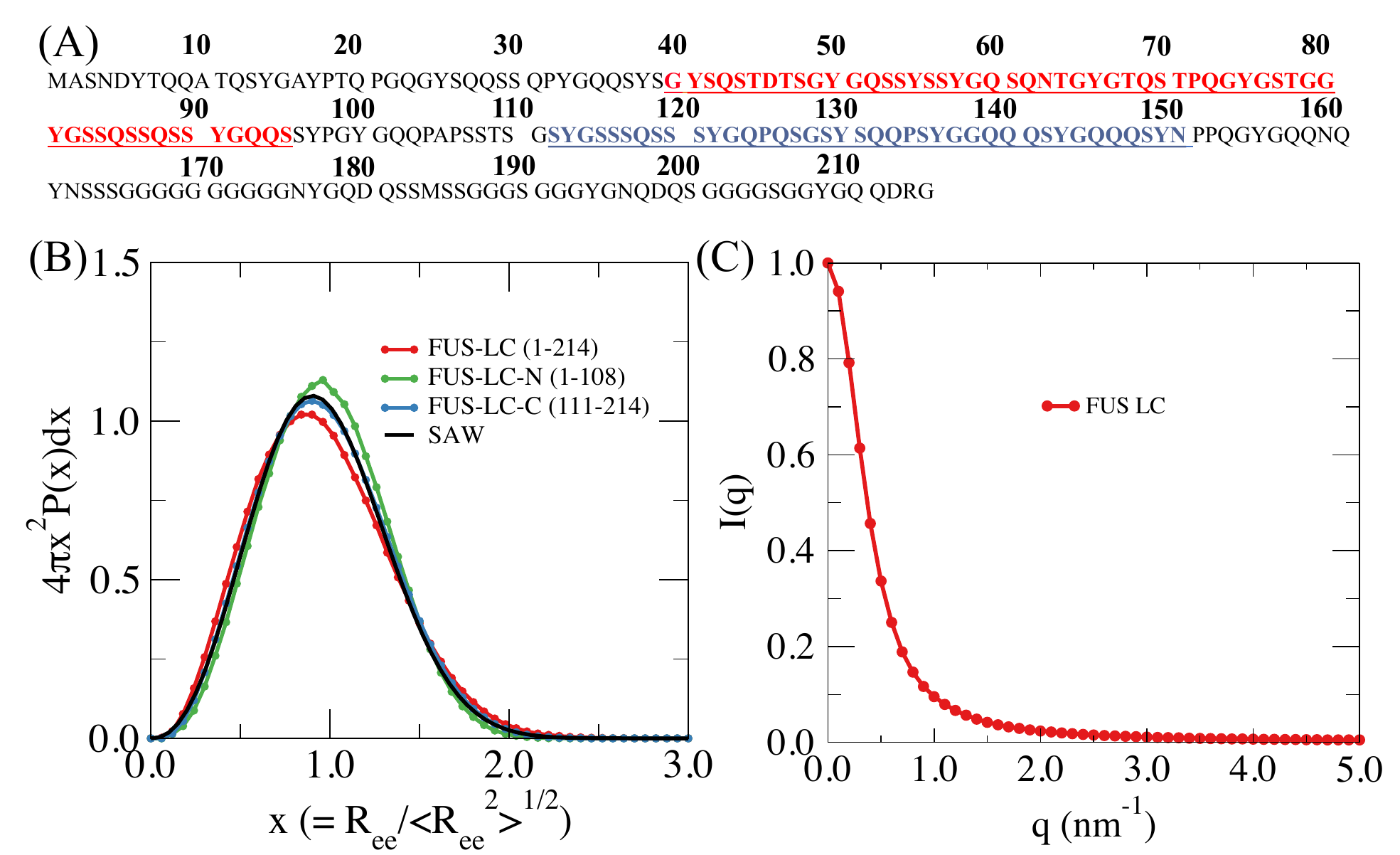}
\caption{(A) Sequence of the Fused in Sarcoma (FUS) Low Complexity (LC) domain.  (B) Distributions of the scaled end-end distance for different FUS-LC variants exhibit the universal Flory random coil behavior. \textcolor{black}{The abbreviation SAW stands for Self-Avoiding walk, an alternate phrase used to describe random coils in the polymer literature.}. (C) The scattering profile for FUS-LC is reminiscent of random coils.  }
\end{center}
\end{figure}

In order to ascertain if FUS-LC and its variants exhibit the characteristics of a {\color{black} random coil}, we  generated a monomer conformation ensemble (MCE) of FUS-LC and truncated versions, FUS-LC-C (residues 110-214) and {\color{black} FUS-LC-N (residues 1-108)} using simulations based on   the SOP-IDP model.~\cite{Kumar21JPCL} Analyses of the MCE shows that both these constructs are devoid of secondary structures, but exhibit the properties of a homopolymer in a good solvent. For a Flory random coil, the distribution of $x = \frac{R_{ee}}{\langle R_{ee} \rangle}$ should exhibit the following universal behavior,
\begin{equation}
\label{dist}
P(x) \sim C x^g exp(-B x^{\delta}),
\end{equation}
\noindent where $B$ and $C$ are constants, the correlation hole exponent, $g \approx 0.28$, and $\delta = \frac{1}{(\nu -1)}$. The distributions $4 \pi x^2P(x)$ for FUS-LC, {\color{black}FUS-LC-N}, and FUS-LC-C obtained from simulations show little deviation from Eq. \ref{dist} (Fig. 1B), thus establishing that the ground states of the three constructs behave as polymers in a good solvent.   Moreover, the calculated structure factor, $I(q)$, has the characteristic behavior of random coils (Fig. 1C).\cite{Kikhney215FEBS} It would be interesting to use to measure $I(q)$ in order to validate our prediction. Thus, we surmise that the ground states of the three constructs behave as polymers in a good solvent, which also supports the conclusions reached for the truncated (residues 1-163) FUS-LC~\cite{Burke15MolCell}.

\begin{figure}[htbp]
\begin{center}
\includegraphics[width=\textwidth]{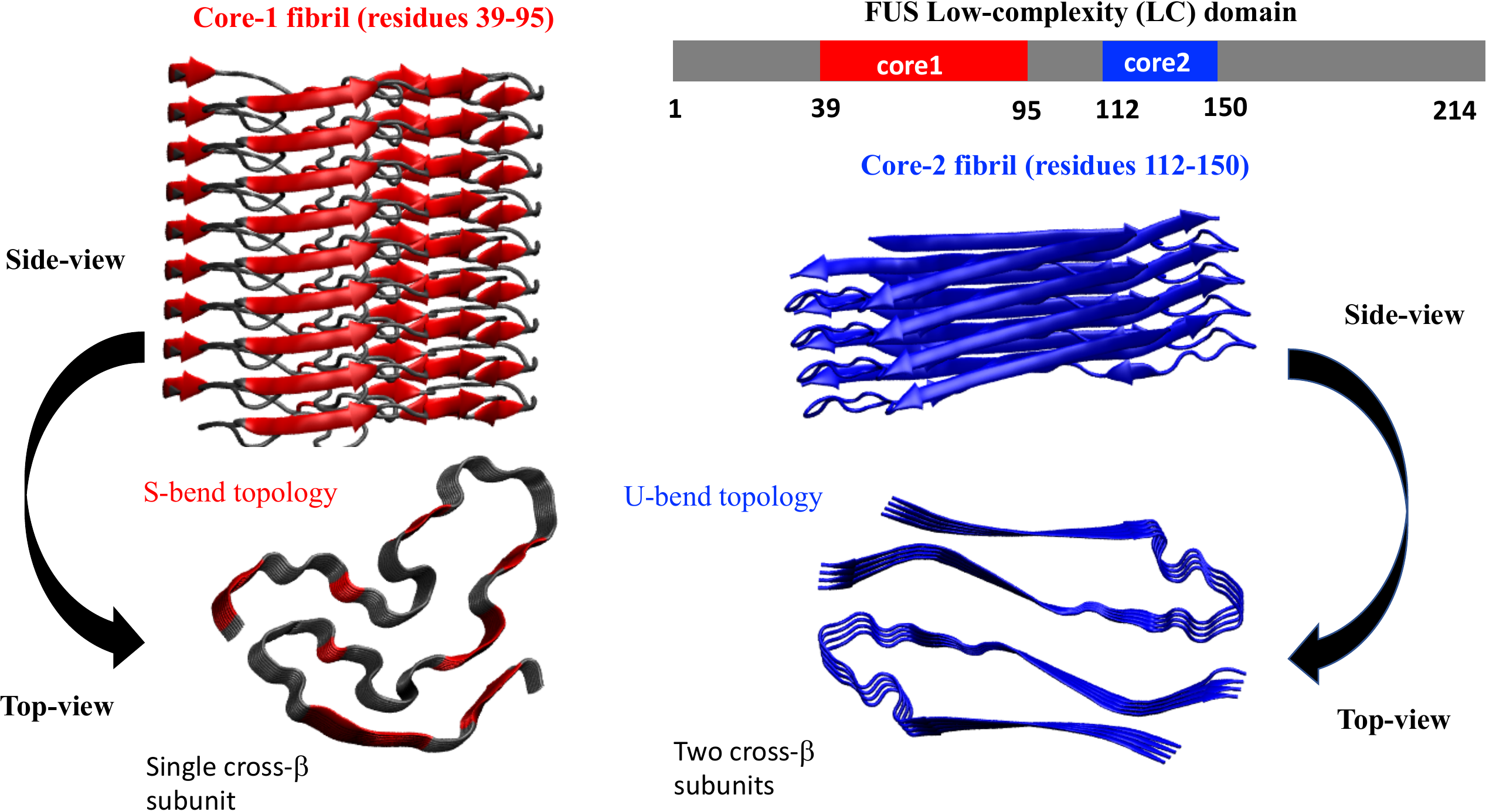}
\caption{Length-dependent structural polymorphism in FUS-LC.   For the full-length FUS-LC,  a S-bend fibril  is formed by residues 39 to 95.\cite{Murray17Cell}  On the other hand,  a truncated variant,  FUS-LC-C (residues 110-124) forms a U-bend fibril \cite{Lee20NatComm},  with each protofilament consisting of two interacting subunits. In this structure,   fibril core is formed by residues 112-150.}
\label{Cores}
\end{center}
\end{figure}

{\bf Length of FUS-LC determines fibril morphology:} In the amyloid and the phase separation fields, there have been attempts to discern the propensity of proteins and IDPs (or IDRs) to self-associate  purely on the basis of sequence (sequence gazing),\cite{TartagliaChemSocRev2008,SerranoJMB2004} an exercise that is not only difficult but also maybe incorrect  because environmental factors (pH, temperature, and presence of {\color{black} crowding agents}) modify the energy landscape substantially.\cite{BrienJACS2012,GieraschJACS2010}  The difficulty maybe illustrated by contrasting the changes in the fibril structures by systematically chopping the sequences in FUS-LC. For example, solid state NMR experiments have shown that the prion domain in Sup35 adopts parallel $\beta$-sheet structures that lies perpendicular to the fibril axis~\cite{Shewmaker06PNAS}, much like many other amyloid fibrils (for example A$\beta_{42}$~\cite{Tycko15Neuron} and mammalian prions see for example~\cite{Cobb07PNAS,Wang20NSMB}). In addition, the structure of the N-domain is not significantly altered in the construct containing both the N-domain  and the highly charged M-domain (residues 124 - 253)~\cite{Shewmaker06PNAS}. 

In sharp contrast, the formation of ordered fibril structures in FUS-LC depends on the length of the  sequence. The 214-residue forms a non-polymorphic fibril (core-1 spanning {\color{black} residues 39-95}) containing a S-bend structure~\cite{Murray17Cell}. The remaining residues are disordered, and flank the two sides forming a fuzzy-coat or brush-like structure.  In FUS-LC-C (residues 110-214)  residues 112-150 also form a non-polymorphic structure but with an entirely different morphology. The protofilament is a dimer with each monomer  resembling a U-bend~\cite{Lee20NatComm}. The dramatic length dependent  variation in the fibril structures of FUS-LC is puzzling, and  lacks a theoretical explanation, although entropic repulsion between the fuzzy coats is likely at play~\cite{Lee20NatComm}.   



 \begin{figure}[htbp]
\begin{center}
\includegraphics[width=\textwidth]{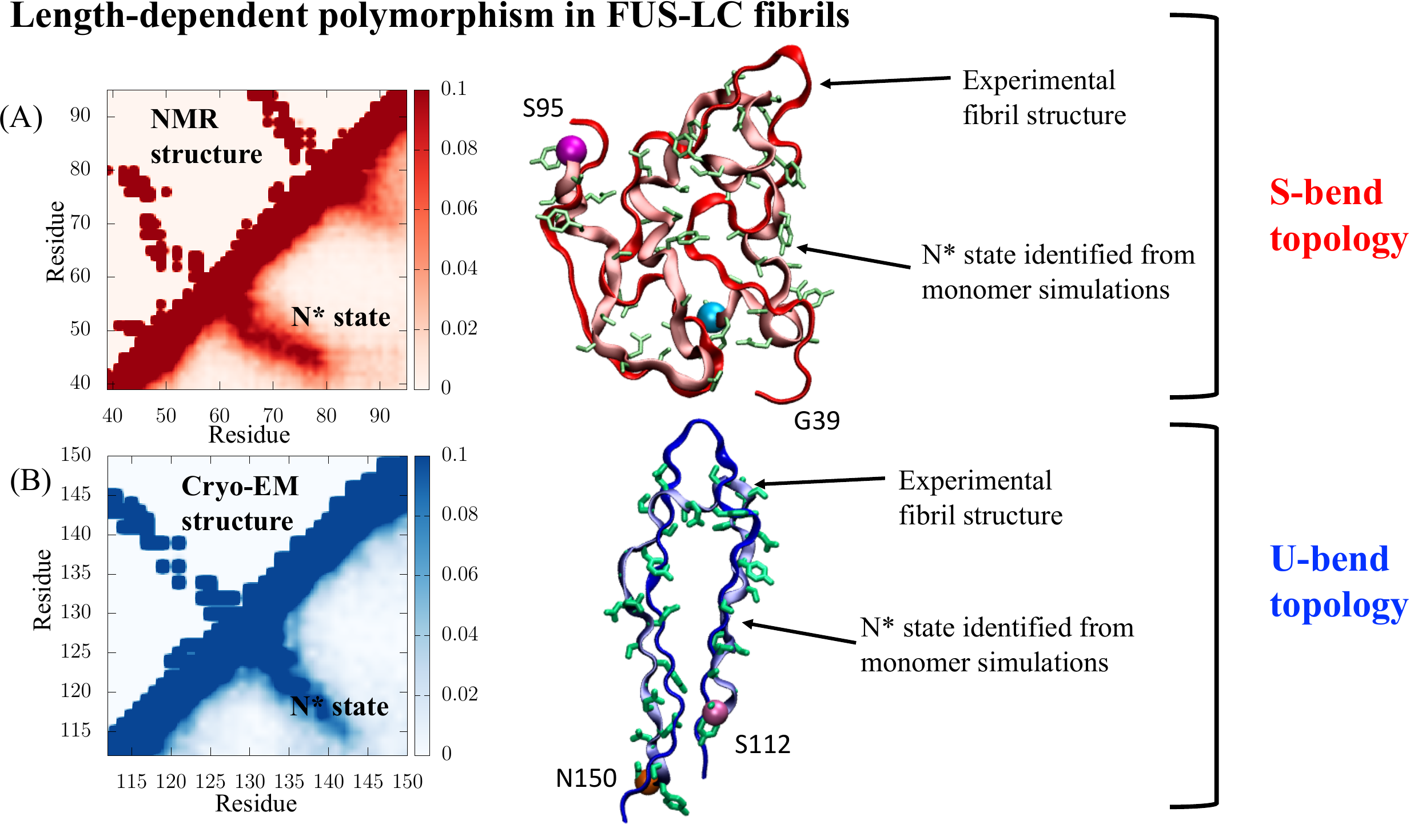}
\caption{Contact maps for the N$^{*}$ ensemble identified from simulations corresponding to the core-1 fibril topology (A) and the core-2 fibril topology (B).    The upper triangles in (A) and (B) were computed from the contacts within the monomer units in the experimental fibril structures.  For illustration,  representative all-atom snapshots from the N$^{*}$ ensembles  are also shown.   It is evident that the N$^{*}$ states bear striking resemblance to the experimental fibril structures.}
\label{Excited}
\end{center}
\end{figure}

{\bf Excited (N$^*$) states  of monomer ensembles have fibril-like structures:} We had shown in a number of publications that amyloid-like structures~\cite{Tarus06JACS,Li10PRL,Chakraborty20PNAS} are the excited states in the MCE.   The excited states N$^*$ states are sparsely populated  (as low at 2\% or less), and at present  can only be detected using relaxation dispersion NMR experiments~\cite{Neudekker12Science,Kotler19PNAS} or potentially using single molecule pulling experiments, as proposed using computations~\cite{Zhuravlev14JMB}.  We showed using simulations~\cite{Kumar21JPCL} that the N$^*$ states, bearing striking resemblances to the S-bend (core-1) and the U-bend (core-2), are populated as excited states in the MCE. We summarize the key results from simulations using a simple model before embarking on the difficult multi-chain problem.

Based on a number of studies in amyloid formation in A$\beta$ peptides~\cite{Tarus06JACS,Chakraborty20PNAS}, we theorized that the entire free energy landscape, especially the lowly populated excited N$^*$ states bearing a strong resemblance to the ordered structures in the fibrils, controls the emergent structural organization and dynamics of the early events in the FUS-LC assembly.  In order to validate this proposal in FUS-LC, let us define the structural overlap parameter~\cite{Klimov98FoldDesign},
\begin{equation} \label{eq1}
\chi(t) = \frac{1}{2(N^2-5N+6)}\sum_{ij}^M H(\Delta - |r_{ij}^{t}-r_{ij}^{fib}|).
\end{equation}
In Eq. 2, the number of residues in the fibril core is $N$, $H(x)$ is the step function, and $r_{ij}^t$ and $r_{ij}^{fib}$ are the pair distances for the simulated conformations and $r_{ij}^{fib}$ is  the corresponding value in the experimental fibril structure  determined for FUS-LC using solid state NMR~\cite{Murray17Cell} or cryo-EM.\cite{Lee20NatComm} We set $\Delta =$ 2 {\AA}, which is a stringent criterion for establishing the similarity between the simulated and experimental fibril-like structures. The total number of pair distances is M = $2(N^2-5N+6)$~\cite{Kumar21JPCL}. Division by $M$ ensures that $\chi$=1 if a conformation coincides with the experimental fibril structure.  Thus, as defined in Eq. \ref{eq1}, the structural overlap satisfies the bound $0 \le \chi \le 1$. 

We calculated $\chi_1$ for core-1 (residues 39-95) and  $\chi_2$ for the 112-150 region using  millions of conformations in the MCE. In so doing, we only considered the simulation coordinates for  the ordered regions. The $\chi$ distribution, $P(\chi)$ FUS-LC  shows that the population of core-1 is marginally higher than core-2 (see Fig. 2 in \cite{Kumar21JPCL}).  Because one might assume that the formation of the more complex core-1 structure (with the S-bend) is less likely  than the simpler U-bend in core-2, a natural expectation  would be that core-2 should have a higher population than core-1. This is essentially an entropic argument but stability is determined by the free energy, which favors core-1, as we describe below. \textcolor{black}{Our assertion is based on a kinetic argument that is based on the assumption that the fluctuations needed to nucleate complex structures from disordered phases should take long time, which indeed would accord well with the Ostwald's rule of stages (see below).\cite{Ostwald}}

 If $\chi$ exceeds a threshold value then the structures resemble the monomer unit from the fibril. The N$^*$ states for core-1 and core-2 superimposed on the fibril-like structures illustrate that this indeed the case Fig. 3, thus validating the proposal that sparsely populated states in the MCE are indicators of the propensity of disordered domains of FUS to form fibrils. The  overall topology of the N$^{*}$ states for both core-1 and core-2 are similar to  the experimental structures (Fig. 3), which is gratifying even though the parameters in the energy function were not adjusted to obtain agreement.  It is worth noting that the structure of the N$^{*}$ state for {\color{black} a truncated variant of FUS-LC (residues 1-163)} is topologically similar to the core-1 fibril structure. Furthermore, the core-1 structure is preserved in FUS-LC-N (residues 1-108),  an observation that is in accord with the findings of Tycko and coworkers.\cite{Lee20NatComm} Taken together, these findings show that as long as the sequence corresponding to core-1 is present, the fibril with S-bend structure is likely to be more probable under nominal experimental conditions, which we explain using free energy estimates.   
 
In sharp contrast to  ssNMR studies~\cite{Murray17Cell,Lee20NatComm}, which show the formation of ordered core-1 (FUS-LC) and core-2 (FUS-LC-C), other NMR experiments were used to suggest that  the droplets in the construct containing residues 1-163 are mobile exhibiting liquid-like behavior~\cite{Burke15MolCell,Fawzi21COSB,Murthy19NSMB}. The experimental conditions in all the quoted experiments are similar. Our computations predict that core-1 should form in the 1-163 FUS construct \cite{Kumar21JPCL}. A possible reason for the discrepancy is that the liquid-like droplets have not aged long enough for fibril formation. It is important to resolve this controversy. Early studies \cite{Patel15Cell} suggested that the full length construct, containing the {\color{black} RNA binding domain}, does aggregate. The structure of the aggregated region, which is envisioned to be solid-like, is unknown. The formation of the solid-like component is accelerated by disease causing mutation, G156E.\cite{Patel15Cell}

{\bf Free energy of core-1 is lower than core-2:} The relative free energies of the core-1 and core-2  structures, calculated using the conformations generated in the monomer simulations, rationalizes the absence of core-2 in FUS-LC. First, we enumerated the number of conformations that belong to core-1 and core-2 using $\chi$ as the order parameter.  From our calculations \cite{Kumar21JPCL} we find that the free energy difference between core-1 and core-2 using $\Delta G = G_{1} - G_{2} = -k_{B}T\times \ln(P_{core-1}/P_{core-2}) = -0.593 \times \ln(8165\pm103/666\pm29)$ = -1.47 $\pm$ 0.005 kcal/mol. The estimates from solubility measurements for $\Delta G$ is even less ($\approx 0.35 \pm 0.19$ kcal/mol). It is interesting that $\Delta G \approx 2.5k_BT$ is sufficient to completely (at least below detectable levels in experiments) eliminate core-2 formation in FUS-LC.  From the calculated energy  and free energy difference, the entropy change maybe readily estimated. We find that  $T\Delta S$ $\approx$ -1.08 $\pm$ 0.38\,kcal/mol and the entropy difference $\Delta S$ is $S_{1} - S_{2}$ = -3.62 $\pm$ 1.27 \,cal/mol/K. Thus, the formation of the S-bend is entropically disfavored relative to the U-bend (the former has a more complex topology than the latter). The unfavorable entropy is compensated by the more favorable energy, leading to the prediction that core-1 is {\color{black} preferred in terms of free energy} over core-2.

{\bf Ostwald rule of stages:} A puzzling feature in the fibril morphology of FUS-LC is that the core-2 structure, detectable only in the absence of the sequence giving rise to core-1, (FUS-LC-C residues 110-214), is completely different from core-1.  A very small sub-population of excited conformations adopts the U-bend-like structure (Fig. 3B) found in experiments~\cite{Lee20NatComm}. More recently, Kato and McKnight have found that in the construct, labeled CTC~\cite{Kato21PNAS}, yet another structure (core-3) emerges, although the structure of core-3 is currently unknown. {\color{black} Kinetic experiments by Kato and Mcknight\cite{Kato21PNAS} illustrate that in FUS-LC core-3 forms faster than core-1. The stability of cores was shown to follow exactly the opposite order, based on reverse-phase column chromatography experiments. In other words, the time to form the cores is inversely correlated with stability.}
The inverse correlation is consistent with the Ostwald's rule of stages,\cite{Ostwald} which we recently showed holds in the formation of distinct polymorphic structures of A$\beta_{42}$ peptides~\cite{Chakraborty23ScienceAdvances}. \textcolor{black}{Ostwald's rule states that if there are two or more polymorphic structures then kinetically the least stable structure would form first. In other words, the rate of formation is inversely correlated with the thermodynamic stabilities of the polymorphic structures.\cite{Ostwald}}

{\bf Effect of phosphorylation - location matters:} Murray et. al.,~\cite{Murray17Cell} systematically examined the effects of phosphorylation by {\color{black} DNA-dependent protein kinase (DNA-PK)} on the stability of the ordered S-bend structure in FUS-LC. Among the 14 phosphorylated  residues, six (S42, S44, S61, T68, S84, and S87) are in core-1. By mutating these sites to Ala, mostly two at a time,  they~\cite{Murray17Cell} discovered that, in all likelyhood,  core-1 fibril structure is destabilized. Phosphorylation within core-1 sites significantly reduced the propensity of FUS-LC monomers to bind to hydrogel droplets, while phosphorylations outside the fibril core had little impact. If Ser and Thr outside the fibril were mutated to Ala the effect was weaker, which suggests that site-specific phosphorylation plays an important role in the stability of the fibril.  A similar conclusion was reached in another experimental study~\cite{Ding20JMB} in which the effect of phosphorylation at S61 in a fragment (residues 50 - 65, referred to as R2)  was assessed. Based on a number of measures, the authors concluded that phosphorylation at S61 abolished both intra and inter molecular interactions needed for aggregation to occur.   

\begin{figure}
\includegraphics[width=0.90\textwidth]{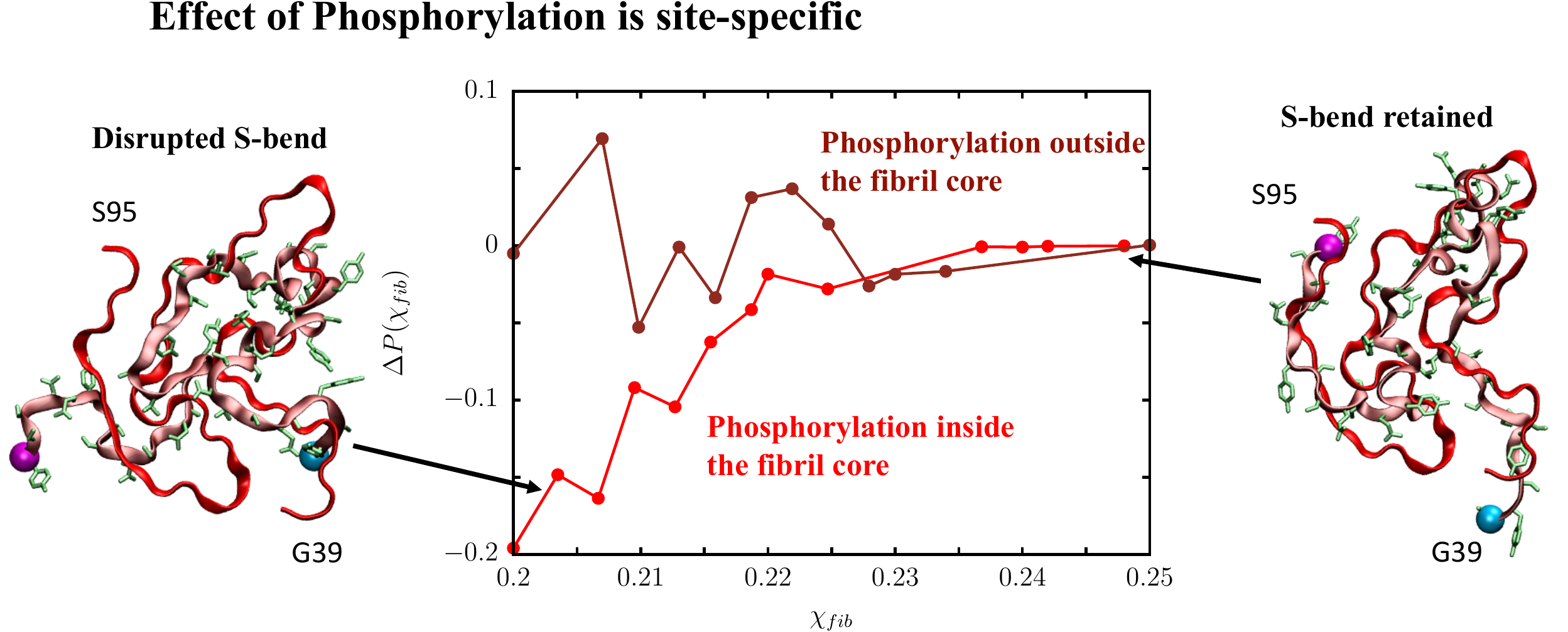}
\caption{The effect of phosphorylation on the MCE of FUS-LC is site-specific. When five Serines and a Theronine within the core-1 fibril are phoshorylated, the key interactions within the N$^{*}$ state are destabilized. None of the structures within the MCE retain the complete S-bend morphology. In contrast, phosphorylation of residues outside the fibril core is less disruptive, and most configurations within MCE retain the S-bend. These trends are reflected in $\Delta P(\chi_{fib})$, which denotes the difference in the probability distribution of $\chi_{fib}$ (similarity with respect to the experimental core-1 fibril structure) between the wildtype and the phosphorylated variants. For clarity, only the tail of the distribution (which corresponds to the N$^{*}$ states within the MCE) is shown. For the P-FUS(in) variant, there is a dip in $\Delta P(\chi_{fib})$, implying that the structural similarity with respect to the fibril state reduces dramatically.}
\end{figure}

In order to model the effects of phosphorylation, we simulated two FUS-LC variants. In one of them, referred to as P-FUS (in), five Serine residues within core-1 (S42, S54, S61, S84, S87) and T68 were phosphorylated by replacing them by the charged  Glu residue, which is clearly a simplification but is often used in computations~\cite{Maciejewski95JBC,Monahan17EMBOJ}. The difference between the distributions, $\Delta P(\chi_{fib}) = P(\chi_{[P-FUS(in)]}) - P(\chi_{[FUS-LC]})$ shows   that the structural similarity of the MCE relative to the fibril state is greatly reduced (Fig.~4). Phosphorylation of sites in the fibril core destabilizes some of the key interactions within the N$^{*}$ state. In particular, for the P-FUS (in) variant,  the conformation is not structurally close to the core-1 fibril  (Fig.~4). The features in $\Delta P(\chi_{fib})$ are less prominent  for the construct P-FUS (out), in which T11, T19, and T30 and S112, S117, and S131 outside of core-1 were phosphorylated. This suggests that the effect of phosphorylation is less disruptive in P-FUS(out) compared to P-FUS (in). Most of the conformations within the N$^{*}$ ensemble seem to retain the overall S-bend topology, which accords well with experimental findings that if the phosphorylation sites are outside the fibril forming region the effects are small.  The conclusion that phosphorylation of residues in the core is destabilizing while the effect is small when residues outside are phosphorylated outside \cite{Murray17Cell} has also been noted in the simulation of droplets. 

The reduced aggregation propensity of the phosphorylated variants could be attributed to the enhanced free energy gap between the disordered ground state and the N$^{*}$ state. Our results suggest that phosphorylation effects are site specific, with disruption occurring if residues in the fibril core are altered either by mutations or phosphorylation. The implication is that specific interactions within and between the FUS-LC polymers stabilize the fibril structure, which was already noted in previous studies~\cite{Murray17Cell,Kato22RNA}.

{\bf FUS aggregates:} The lessons one can learn from careful characterization of monomers are limited.  At finite concentration of FUS polymers many phases could emerge. For instance, FUS-LC (or FUS) could undergo phase separation into liquid droplets, or form fibrils and gels. The droplet could age to a fibril or a glass, depending on external conditions.\cite{KatoCell2012,Murray17Cell,WangCell2018,JawerthScience2020} Both fibrils, and glasses are solids (they resist shear). We use the term fibril here to describe the ordered structures found in solid state NMR experiments \cite{Murray17Cell,Lee20NatComm} or more recently by cryo-EM \cite{Sun22iScience}. Computer simulations~\cite{Garaizar22PNAS}, using a single-bead per residue coarse grained model~\cite{Dignon18PLOSCompBiol}, show that as the droplet ages  the morphologies of FUS could change, producing  non-equilibrium structures. Some of the scenarios that could arise are schematically shown in Fig.~5. At a fixed temperature, various phases of FUS (for that matter any protein or IDP) could emerge as external conditions (for example, salt concentration, crowding agents and pH) are altered. In other words, the phase diagram for FUS-LC or its variants is complicated. A survey of experiments in Table 1 and Fig.~5 points to several possibilities.  Here, we only discuss phase separation and fibril formation.

 \begin{figure}[htbp]
\begin{center}
\includegraphics[width=0.90\textwidth]{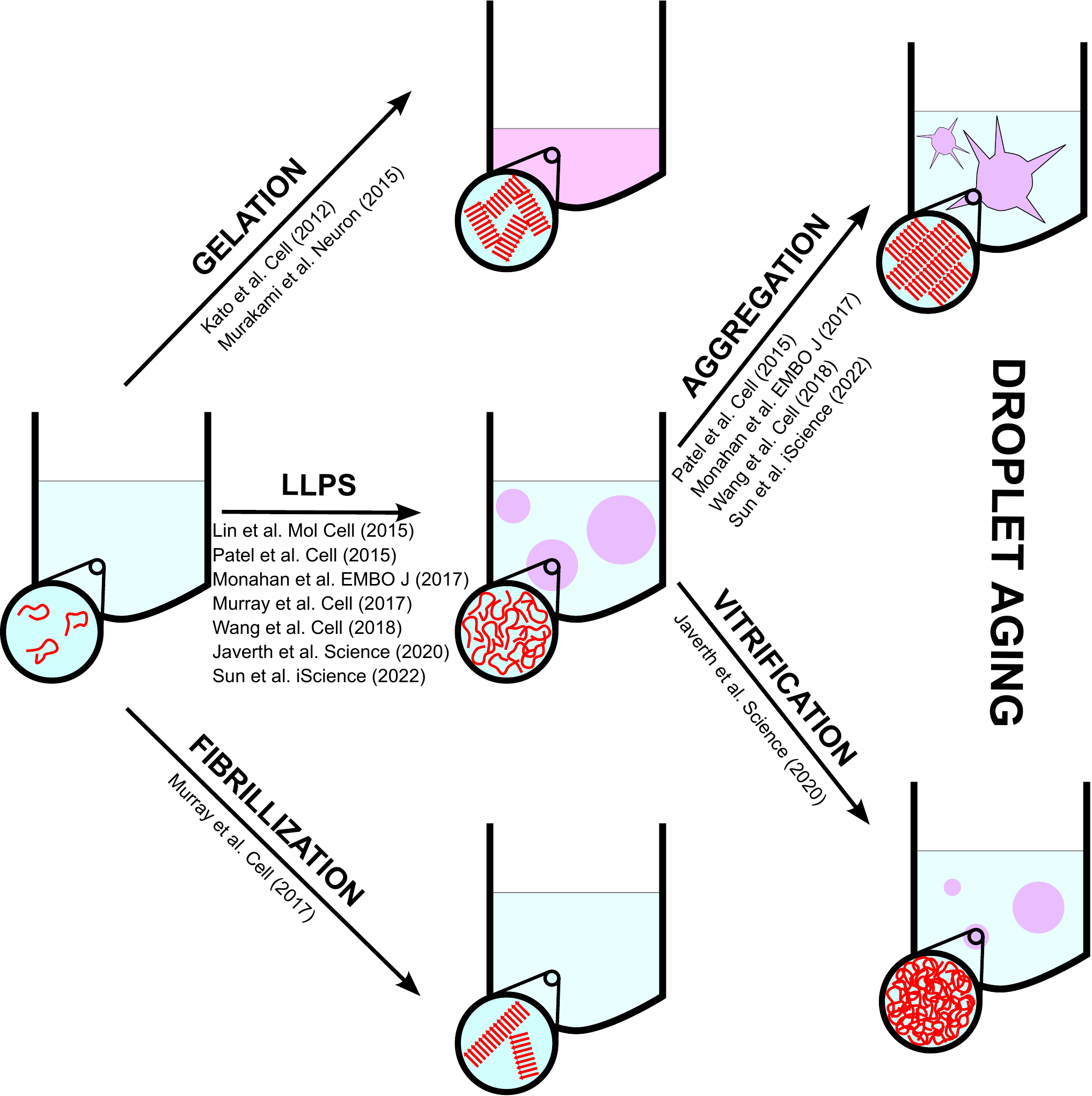}
\caption{A schematic illustrating liquid-solid phase transition in FUS.   The liquid droplet state is metastable,  and ages with time to form hydrogels or irregularly shaped solid aggregates,  which have a characteristic cross-$\beta$ architecture typical of amyloid fibrils.   Recent work also suggests that aging could induce Maxwell glass-like dynamic behavior.\cite{JawerthScience2020}}
\label{Droplet}
\end{center}
\end{figure}

\begin{table}[htp]
\begin{tabular} {| l | l | l | l |  l | l |}
\hline
\small Experimental Reference & \small Sequence & \small [FUS]  & \small [salt] & \small Temp. & \small Observed Phase \\
             &            & \small ($\mu$M)      &   \small ($mM$) & \small (K) &  \\
\hline
\small Kato \textit{et al.}\cite{KatoCell2012} & \small Tag (2-214) & \small 2000  & \small 200 &  \small 277 & \small Gel  \\
\hline
\small Lin \textit{et al.}\cite{LinMolCell2015} & \small SNAP (1-237) & \small 10.75  & \small 37.5 & \small 298 & \small Liquid droplets  \\
\hline
\small Patel \textit{et al.}\cite{Patel15Cell} & \small FUS GFP & \small 10 &  \small 500 & \small 298 & \small Fibrils \\
         & \small FUS GFP & \small 500 & \small 200 &  \small 277 & \small Gel\\
 \hline
\small Murakami \textit{et al.}\cite{MurakamiNeuron2015} & \small FUS-LC(2-214) & \small 1000 & \small 500 & \small 296 & \small Hydrogel \\
    &  \small FUS-LC+RBD & \small 1-5 & \small 150 &  \small 277 & \small Liquid droplets\\
  \hline
\small Murray \textit{et al.}\cite{Murray17Cell} & \small FUS-LC(1-214) & \small 30 & \small 200 & \small 297 &  \small Fibril \\
\hline
\small Monahan \textit{et al.}\cite{Monahan17EMBOJ} & \small FUS-LC(1-163) &  \small 5 &  & \small 298 & \small Liquid droplets\\
        & \small FUS-LC(1-214) & \small 5 & \small 150 & \small 298 & \small Liquid droplets\\
\hline
\small Wang \textit{et al.}\cite{WangCell2018} & \small FUS-LC+RBD  & 5 & 150 & & Liquid droplets \\
 & \small FUS-LC(1-214) &  \small > \small 10 & \small 150 & & \small Liquid droplets\\
 \hline
\small Jawerth \textit{et al.}\cite{JawerthScience2020} & \small FUS-LC+RBD & 10 & 75 & & \small Maxwell glass \\
\hline
\small Berkeley \textit{et al.}\cite{Berkeley21BJ} & \small 5\%Cy3- FUS(1-163) & & \small 150 &  \small 298 &  \small Fibril\\
\hline
\small Sun \textit{et al.}\cite{Sun22iScience} & \small mCeru-FUS-LC(2-214) & \small 100 & \small 100 & \small 310 & \small Fibril \\
\hline
\end{tabular}
\caption{A list of previous work that have experimentally characterized various phases for the FUS protein.  For clarity,  the precise experimental conditions,  including the inclusion of tags,  protein and salt concentration,  as well as temperatures,   are also summarized.}  
\end{table}

\noindent Under certain conditions FUS undergoes phase separation at high polymer concentration. In such a phase, a dense condensate (or coarcervate, term first used in 1929 in the same sense as condensate is used today) coexists with a low density dilute phase. Such a generic phase behavior in FUS (and other IDPs) is explained qualitatively by the venerable Flory-Huggins (FH) lattice theory\cite{FloryBook,FloryJCP1942,HugginsJCP1942} that has only the $\chi_{FH}$, a dimensionless  parameter, which describes the differential interaction energy (relative to $k_BT$) between the monomer and solvent with respect to an average of monomer-monomer and solvent-solvent interactions.  Although FH is a mean-field theory, it is often remarkably accurate if the number of lattice sites ($N$) is large ($N \gg 1$) because the mixing entropy associated with the polymer scales as $\frac{1}{N}$. Note that $N = N_1 + nN_2$ where $N_1$ is the number of solvent molecules, $N_2$ is the number of polymer chains, and $n$ is the degree of polymerization, which in FUS-LC is 214. These quantities are easily expressible in terms of number density or molar concentrations. Interestingly, the FH theory describes phase separation in IDPs  at least qualitatively (for a lucid review see~\cite{Zhou18TIBS}), even though IDPs are not large (in the sense implied in polymer theories).  Because IDPs are heteropolymers, which suggests that the sequence, not merely the overall length, could influence the stability of various structures, as has been shown in experiments involving site-specific phosphorylation.

We focus on two puzzling aspects that pertain to molecular structure of FUS upon aggregation. The survey in Table 1, accompanied by a study of the listed papers, shows that there is a lack of consensus among many different experiments. Some of these are expected because the experimental conditions (concentrations of FUS and salt or pH), and possibly even sequences (FUS-LC and full length FUS that also contains the RNA binding domain),  and the method of preparation are not exactly similar.  There are two caveats: (1) The pH used in most of the studies in Table 1 changes over a narrow range  (between 7.4 and 7.5). (2) We assume that the ground state should be independent of the preparation method unless FUS aggregates are glass-like that are trapped in one of the many metastable states. For now, we disregard this possibility.

{\bf Tyrosine ladder:} We first describe the salient features of core-1 and core-2, shown in Fig.~6. A few observations are worth making: (1) It is self-evident that the aromatic rings in Tyr are beautifully (seemingly in perfect registry) stacked on top of each other in both core-1 and core-2. There are 8 aromatic ladders in core-1. Because core-2 protofilament is a dimer, there are 6 Tyr ladders per chain.  Naturally, the number of Tyr ladders is exactly equal to the number of Tyr in the two cores.  (2) The structures reveal the importance of $\pi-\pi$ interactions  in stabilizing the two cores. That $\pi-\pi$  interactions play an important role in materials with aromatic groups is hardly a surprise given the preponderance of such interactions in graphene,  organic compounds, and base stacking in nucleic acids.\cite{Desiraju1989,PetskoScience1985,Zaric2020,Otyepka2013}

 \begin{figure}[htbp]
\begin{center}
\includegraphics[width=\textwidth]{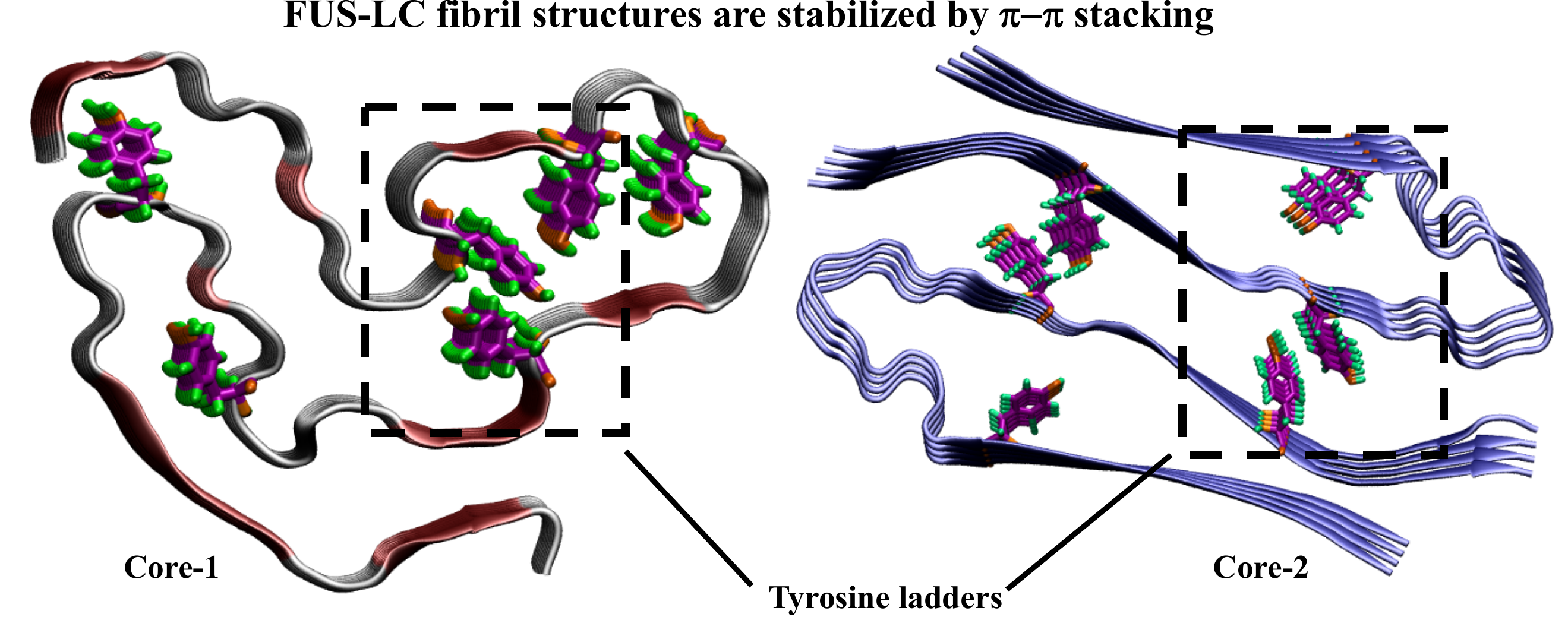}
\caption{Stacking of aromatic Tyrosine side-chains in the core-1 (A) and core-2 fibril (B).  The FUS fibril-cores lack the strong hydrophobic interactions found in other fibrils,  such as A$\beta$ and $\alpha$-synuclein.   It is likely that the Tyrosines stacked in near-perfect registry (forming Tyrosine ladders) stabilize the fibril cores formed by FUS-LC.}
\label{tyr_stack}
\end{center}
\end{figure}

The importance of $\pi-\pi$ interactions has been pointed out in the IDP literature (for an excellent review see~\cite{Gomes19JBC}, although it was known in the context of oligomer formation in A$\beta$ peptides. Using molecular dynamics simulations, Klimov and Thirumalai~\cite{Klimov03Structure} showed that the hydrophobic core (KLVFFAE spanning residues 16-22 in A$\beta$), which adopts an antiparallel fibril  structure~\cite{Balbach00Biochemistry}, is stabilized by $\pi-\pi$ interaction (in this case between two Phe residues). Mutation of Phe to Ser disrupts the stability of A$\beta$ oligomers~\cite{Klimov03Structure}. A dramatic demonstration of $\pi-\pi$ interactions is found in the self-association of a single amino acid, Phe~\cite{Adler-Abramovich12NatChemBiol}, which is associated with the autosomal disease, phenyletonuria.  It was shown that Phe  forms amyloid-like fibrils, almost entirely stabilized by $\pi-\pi$ interactions.  In the context of FUS, the role of tyrosine in mediating multivalent interactions within condensates has been highlighted recently.\cite{WangCell2018,MartinScience2020} (3) There is only charged residue (Asp46) in core-1 and none in core-2.  It was noted~\cite{Murray17Cell} that Asp46 is either exposed to the solvent or interacts with or Ser44 and Gln52, which are polar. Thus, one can rule out cation-$\pi$ interactions as the source of stability in core-1 and core-2, although it is suspected to play a role in the self-association of FUS that also includes the RNA binding domain. (4)  Hughes et. al.,~\cite{Hughes18Science} determined atomic structures of three peptide fragments in core-1 using X-ray crystallography. The peptides crystallized as pairs of kinked $\beta$-sheets, much like core-2. The structure of residues 54-61 ($^{54}$SYSSYGQS$^{61}$) containing two Tyr forms exactly two aromatic ladders (Figure 1c in ref. \cite{Hughes18Science}). The region, $^37SYSGYS$ also forms fibril-like structure with Tyr stacks (S37 and Y38 are outside core-1). The kinks in these structures arise possibly because the Tyr residues are separated by only two residues (less than the persistence length of peptides). Consequently, there has to be  a kink to accommodate the Tyr ladders. In contrast,  in both core-1 and core-2 the Tyr residues are separated by 5 or more residues. (5) Superposition of the X-ray structure for this region with the solid-state NMR structure  shows significant deviations. However,   the fragments and the core-1 structures both contain the tyrosine ladder, that must stabilize the ordered structures. In FUS-LC, the favorable inter peptide region in the ordered region (core-1) must compensate for the entropic repulsion between the fuzzy disordered coats that flank the S-bend. The lengths of the two coats span residues 1-38 and residues 96-214.  {\color{black} More importantly,  it is not merely the patterns of sequences or their overall composition, but the location of key residues that determine the propensity of IDPs to self-associate.}

\section{Discussion}

Although there are several common global features that FUS-LC shares with other monomeric IDPs, it also exhibits certain properties at finite concentrations that are most intriguing. With the current perspective as a background, there are several open questions that we believe are not satisfactorily answered. In what follows we list them in no particular order.

\begin{itemize}

\item From a theoretical perspective, it is puzzling that the formation of  non-polymorphic fibrils depends on the length of the FUS-LC.  Notwithstanding reports \cite{Burke15MolCell}, which assert that aggregates consisting of residues 1-163 are disordered as are the monomers,  studies have shown that fibrils indeed form in distinct regions. Why do core-1 and core-2 not coexist? The answer may well be due to entropic repulsion between the fuzzy coats outside the ordered region, but this picture has to be quantified.  To make matters interesting there is a hint that a third core-3 structure (as yet undetermined) emerges in a construct with residues beyond 150.\cite{Kato21PNAS}

\item To make matters interesting, a new serpentine type fold comprising of nine $\beta$ strands, with order in residues between 34-124, has been determined using cryo-EM \cite{Sun22iScience}. Explanation of the stability of this fourth structure, which contains residues in core-1 and parts of core-2, is lacking. The serpentine fold is also suspected to be non-polymorphic.

\item Why are FUS-LC fibrils non-polymorphic whereas A$\beta$ fibrils are not? It is argued that the primary source of stability of A$\beta$ is the inter-peptide hydrophobic interactions. On the other hand, it is suspected that FUS-LC fibrils are stabilized predominantly by hydrophilic interactions. The N domain of yeast prions, which forms fibrils with cross-$\beta$ structures that are likely stabilized by hydrophilic interactions, may exhibit polymorphism. What are the general (if there are any) sequence-dependent characteristics that determine the origin of  polymorphic structures?  We hasten to add that  even in FUS-LC, mutation of Tyr (in the Tyr ladder) is likely to disrupt the stability of the fibril. 

\item Aggregation rate in the disease mutation, G156E, in FUS with or without the RNA binding domain, is enhanced relative to the wild type (WT) \cite{Patel15Cell,Berkeley21BJ}. It is worth noting that the fibril formation rate in the Dutch mutant, E22Q, in  A$\beta$ peptides increases compared to the WT. Decrease in the total charge in the Dutch mutant was used to rationalize the enhancement in the rate of fibril formation \cite{Massi02ProtSci}. A similar reasoning does not explain the enhancement in the aggregation rate in the G156E mutant. In the G156E mutant, which apparently has  the same fibril morphology as the WT \cite{Berkeley21BJ}, there is an increase in charge, which would enhance electrostatic repulsion. Furthermore, association between different molecules could require, at least, partial desolvation of Glu, another unfavorable process. Thus naively, one would expect that fibril formation rate in the G156E mutant should be less than in the WT. What are the factors that could quantitatively rationalize the experimental observation?

\item In a most insightful paper it was shown, using passive and active rheology, that IDP/IDR condensates age with time, much like spin glasses \cite{Bouchaud92JPhysI} where aging effects have been examined using the trap model.  The full length 526-residue FUS, which has no discernible fibril-like order in the interior of the condensate, resembles an amorphous material. Surprisingly, a single relaxation time ($\tau_c$), which increases with the waiting time ($t_w$) dramatically, characterizes the dynamics of the condensates which undergoes no significant structural change over time \cite{JawerthScience2020}. By analogy with Maxwell fluids, which describes viscoelastic relaxation using a single time constant,  It was argued that protein condensates are like a Maxwell glass.   There are two reasons that this result is unexpected. (i) From the perspective of the dynamics in supercooled liquids  one expects a spectrum of relaxation times, often expressed in terms of a stretched exponential function \cite{Berthier11RMP,Kirkpatrick15RMP}. (ii) The energy landscape of IDPs and IDR is rugged containing multiple minima, separated by free energy barriers \cite{Chebaro15SciRep}. One would, therefore, expect that transitions from between the minima would lead to multiple relaxation times, which would be incompatible with the Maxwell glass picture. In such a scenario, it appears that protein condensates  explore only a single rough minimum, but with minima within minima (hierarchical structure) \cite{Zwanzig88PNAS}. The free energy barriers between the sub-minima  increase as the system relaxes to the bottom of the minimum, thus accounting for increase of $\tau_c$ with $t_w$.   A molecular explanation of this  finding, which appears to be general \cite{Jawerth20Science} in interacting IDPs, would be most interesting.    

\end{itemize}

\section{Conclusion}

Although we have focussed on the characteristics of FUS-LC and variants, it is likely that the peculiar features, such as the length-dependent changes in fibril morphology, responses to perturbations (phosphorylation), non-polymorphic nature of the fibrils may be shared with TDP43 \cite{Fonda21JACS}. Exploring the generic sequence features that translate into the distinct condensate morphologies, which apparently are very sensitive to external conditions, remains a major problem. Finally, what are the common themes in the fibril formation and growth of IDPs, such as A$\beta$ and $\alpha$-synuclein, which result in polymorphic structures and low complexity IDP sequences? Answering this question and others raised here should go a long way towards a general theory of condensate formation.  

\newpage

\acknowledgement {\color{black} We are grateful to Rob Tycko for discussions on topics related to FUS and to J. P. Bouchaud on the trap model.} We also acknowledge comments by Ryota Takaki. We acknowledge the Texas Advanced Computing Centre (TACC) for providing computing resources.  Our work was supported by grants from the National Institutes of Health (GM-107703), National Science Foundation (CHE 19-00033), as well as a grant from the Welch Foundation (F-0019) administered through the Collie-Welch Regents Chair.  A part of this work was initiated when MLM was a research associate at The University of Texas at Austin. 

\medskip

\noindent The authors declare no competing financial interest.
\clearpage

\bibliography{IDP.bib}

\end{document}